\documentclass[12pt]{revtex4}
\usepackage{amssymb}

\usepackage{amsmath}
\usepackage{graphicx}

\begin{document}

\title{Statistical Assemblies of Particles with Spin}
\author{G. Ramachandran}
\affiliation{G. V. K. Academy, Jayanagar, Bangalore - 560070, India}
\email{gwrvrm@yahoo.com;  gvkacademy@gmail.com}
\maketitle

\hspace{2.66in}{\bf{Abstract}}

Spin, $s$ in quantum theory can assume only half odd integer or integer values. For a given $s$, there exist $n=2s+1$ states %%@
$|s,m\rangle$, $m=s,s-1,........,-s$. A statistical assembly of particles (like a beam or target employed in experiments in %%@
physics) with the lowest value of spin $s=\frac {1}{2}$ can be described in terms of probabilities $p_m$ assigned to the two %%@
states $m=\pm \frac {1}{2}$. A generalization of this concept to higher spins $s>\frac {1}{2}$ leads only to a particularly simple %%@
category of statistical assemblies known as `Oriented systems'. To provide a comprehensive description of all realizable %%@
categories of statistical assemblies in experiments, it is advantageous to employ the generators of the Lie group $SU(n)$. The %%@
probability domain then gets identified to the interior of regular polyhedra in $\Re^{n-1},$ where the centre corresponds to an %%@
unpolarized assembly and the vertices represent `pure' states. All the other interior points correspond to `mixed' states. The %%@
higher spin system has embedded within itself a set of $s(2s+1)$ independent axes, which are determinable empirically. Only when %%@
all these axes turn out to be collinear, the simple category of `Oriented systems' is realized, where probabilities $p_m$ are %%@
assigned to the states $|s,m\rangle$. The  simplest case of higher spin $s=1$ provides an illustrative example, where  additional %%@
features of  `aligned'  and  more general `non oriented'  categories are displayed . \vspace{0.7in}

\vspace{1cm}

\newpage 

\vspace{2cm}

\begin{center}
 \vspace{2cm}

	\textbf{Dedicated with gratitude to\\ Dr. C. R. Rao \\ on the Happy Occasion of his \\ $100^{th}$ Birthday}
\end{center}

\vspace{6cm}

 \begin{center}
\textbf{With due apologies to the great Telugu poet Bammera Pothana}
\end{center}

\begin{figure}[h]
\includegraphics[scale = 0.5]{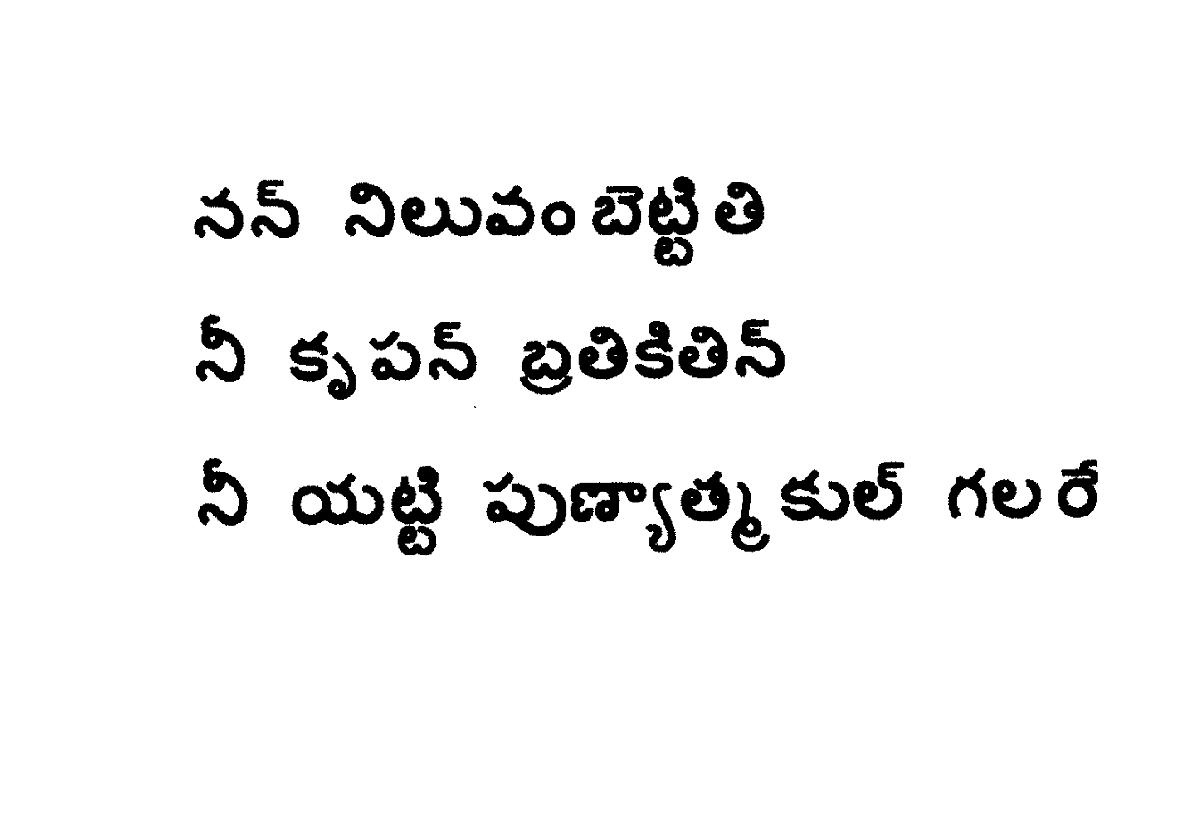}
\end{figure}

\newpage
\begin{section}{Introduction}

`Spin' brings immediately to mind an object rotating about itself. For eg, the Earth, a spinning top etc. When the spin of the %%@
electron was discovered in 1925 by Uhlenbeck and Goudsmit \cite{a}, two great physicists viz., H A Lorentz (Nobel Laureate, 1902) %%@
and Wolfgang Pauli (Nobel Laureate, 1945), who opposed the concept of electron spin,  were proved wrong as they had in mind a %%@
classical electron with an estimated radius~$2.8\times 10^{-13}cm$. To the best of experimental information to date, the electron %%@
does not appear to have any size!\\
`Spin' is an intrinsic attribute like mass or electric charge for a particle. Atomic nuclei also have 'spin', which is essentially %%@
the sum total of the intrinsic spins of the nucleons and their orbital angular momenta. The quantum of radiation, viz., the photon %%@
also has  'spin', although it has no mass. In fact, the concept of polarization was first introduced in the context of light.  %%@
Beams or targets which are essentially statistical assemblies of particles/nuclei with polarized spin are employed in several %%@
experiments in physics. We are considering here such statistical assemblies  The language used is mostly that of physics due to  %%@
my lack of knowledge of mathematical terminology employed by experts on probablity theory.\\
\end{section}

\begin{section}{Quantum Description of Spin}

In the quantum description of the submicroscopic world, any observable $A$ associated with a physical system is represented by a %%@
bounded linear operator $A$ in a separable Hilbert space ${\cal H}$, which accomodates all possible states $|\psi\rangle$ of the %%@
system as vectors in ${\cal H}$.
According to the Ehrenfest theorem \cite{b}, the quantum expectation value
\begin{equation}
\langle A \rangle\equiv\langle\psi|A|\psi\rangle,
\label{a}
\end{equation}
represents the classical value of the observable $A$, which indeed satisfies classical laws.\\
Spin $\vec{S}$, in quantum theory, is defined through the commutation relationships
\begin{equation}
\vec{S}\times\vec{S}=i\hbar\vec{S}
\label{b}
\end{equation}
satisfied by its components, $S_{x}=-i\hbar\,G_{yz},$$S_{y}=-i\hbar\,G_{zx}$, $S_{z}=-i\hbar\,G_{xy}$, where $G_{xy}$, $G_{yz}$ %%@
and $G_{zx}$ are the generators of the Lie group SO(3) of rotations in the $x-y$, $y-z$ and $z-x$ planes respectively of the real %%@
three dimensional physical space. The Lie algebra of this group admits of matrix representations in terms of $n\times n$ matrices, %%@
where $n$ is related to the spin, $s$ of a particle through
\begin{equation}
n=2s+1
\label{c}
\end{equation}
The electron has spin $s=\frac{1}{2}$. The spin hypothesis \cite{a}, in the words of Condon and Shortley \cite{c},``quickly %%@
cleared up many difficult points, so that it at once gained acceptance". We use the universally adopted Condon and Shortley phase %%@
conventions \cite{c}. We also employ the natural system of units; 
\begin{equation}
\hbar=1,\,c=1
\label{d}
\end{equation}
where $\hbar$ symbolizes $\left(2\pi\right)^{-1}$ times the Planck constant $h$ and $c$ denotes the velocity of light. 
The group SO(3) is homomorphic to the group SU(2) of unitary unimodular $2\times2$ matrices. For $s=\frac{1}{2}$
\begin{equation}
\vec{S}=\frac{1}{2}\,\vec{\sigma},
\label{e}
\end{equation}
where the three components of $\vec{\sigma}$ viz.,
\begin{equation}
\sigma_{x}=\left(
\begin{array}{cc}
	0&1\\
	1&0
\end{array}\right);\,
\sigma_{y}=\left(
\begin{array}{cc}
	0&-i\\
	i&0
\end{array}\right);\,
\sigma_{z}=\left(
\begin{array}{cc}
	1&0\\
	0&-1
\end{array}\right)
\label{f}
\end{equation}
are referred to as the Pauli spin matrices.
\end{section}
\begin{section}{Spin States}
In general, for any given spin value $s=\frac{1}{2},1,\frac{3}{2},2,....,$there exist a complete orthonormal set of $n$ spin %%@
states $|s,m\rangle,$ $m=s,s-1,....,-s$, which are simultaneous eigen states of $S^2=\left(\vec{S}.\vec{S}\right)$ and $S_z$ such %%@
that 
\begin{equation}
S^2|s,m\rangle = s(s+1)|s,m\rangle,
\label{g}
\end{equation}
\begin{equation}
S_z|s,m\rangle = m|s,m\rangle,
\label{h}
\end{equation}
\begin{equation}
(S_x\pm iS_y)|s,m\rangle=\sqrt{(s\mp m)(s\pm m+1)}|s,m\pm1\rangle,
\label{i}
\end{equation}
if we choose the axis of quantization to be along the $z$-axis of a right handed cartesian co-ordinate system.\\
In the particular case of $s=\frac{1}{2}$ and $n=2$, the two states 
\begin{equation}
|s=\frac{1}{2}\;,\,m=+{\frac{1}{2}}\rangle=
\left(
\begin{array}{cc}
	1\\
	0
\end{array}\right)=\uparrow;   \
|s=\frac{1}{2}\;,\,m=-{\frac{1}{2}}\rangle=
\left(
\begin{array}{cc}
	0\\
	1
\end{array}\right)=\downarrow
\label{j}
\end{equation}
are referred to as `up' and `down' spin states or spinors. They readily satisfy  (\ref{g}) to (\ref{i}), as can be seen using %%@
(\ref{e}) and (\ref{f}). We refer to all values $s\geq$ as higher spins.
\end{section}

\begin{section}{Density Matrix in Quantum Theory}
The elegant concept of the density matrix $\rho$ was introduced by von Neumann and others \cite{d}, which is ideally suited for %%@
our purpose. Any quantum state $|\psi\rangle$, which is a vector in a Hilbert space ${\cal H}$, may be expressed as 
\begin{equation}
|\psi\rangle=\sum_{i}C_i|\varphi_i\rangle,
\label{k}
\end{equation}
in terms of a conveniently chosen complete orthonormal set of basis states $\{|\varphi_i\rangle\}$ in ${\cal H}$. The complex %%@
numbers
\begin{equation}
C_i=\langle\varphi_i|\psi\rangle,\;C^{*}_{i}=\langle\psi|\varphi_i\rangle
\label{l}
\end{equation}
are complex conjugates of each other and 
\begin{equation}
\sum_{i}{\left|C_i\right|}^{2}=1,
\label{m}
\end{equation}
for normalized states, $\langle\psi|\psi\rangle=1.$ 
If,
\begin{equation}
A_{ji}=\langle\varphi_j|A|\varphi_i\rangle
\label{n}
\end{equation}
denote matrix elements of a bounded linear operator $A$ in ${\cal H}$ with respect to the chosen basis, the quantum expectation %%@
value (\ref{a}) of the observable $A$ may be expressed as
\begin{equation}
\langle A\rangle=\langle\psi|A|\psi\rangle=\sum_{i,j}\rho_{ij}A_{ji}=Tr(\rho A)
\label{o}
\end{equation}
where $Tr$ denotes trace or spur, while the density matrix $\rho$ is defined through its elements
\begin{equation}
\rho_{ij}=C_{i}C^{*}_{j}=\langle\varphi_{i}|\psi\rangle\langle\psi|\varphi_{j}\rangle
\label{p}
\end{equation}
or
\begin{equation}
\rho=|\psi\rangle\langle\psi|
\label{q}
\end{equation}
which is hermitian i.e.,
\begin{equation}
\rho^{\dagger}=\rho
\label{r}
\end{equation}
The quantum state $\psi$ of the system is thus represented by $\rho$.
Moreover,
\begin{equation}
Tr\,{\rho}=1,
\label{s}
\end{equation}
which is the same as (\ref{m}) and 
\begin{equation}
\rho^{2}=\rho.
\label{t}
\end{equation}

\end{section}

\begin{section}{Spin Density Matrix as a vector in $n^{2}$ Dimensional Linear Vector Space}

Identifying the basis states $\{|\phi_{i}\rangle\}$ with the states $\{|sm\rangle\}$ for a particle with spin $s$, we readily see %%@
that the spin density matrix $\rho$ is a hermitian $n\times n$ matrix. In general, any complex $n\times n$ matrix may be looked %%@
upon as a vector in $n^{2}$ dimensional complex linear vector space of $n\times n$ matrices. If $A$ and $B$ are any two matrices, %%@
their inner product $(A,B)$ may be defined through
\begin{equation}
(A,B)=Tr(AB^{\dagger})
\label{u}
\end{equation}
so that we may conveniently choose an orthogonal basis $\{\Omega_{0}=1,\Omega_{1},...\Omega_{n^{2}-1}\}$ satisfying
\begin{equation}
Tr(\Omega_{i}\Omega^{\dagger}_{j})=n\,\delta_{ij}
\label{v}
\end{equation}
and express $\rho$ in terms of them as
\begin{equation}
\rho=\sum^{n^{2}-1}_{i=0}\omega_{i}\Omega_{i},
\label{w}
\end{equation}
where the coefficients
\begin{equation}
\omega_{i}=\frac{1}{n}Tr(\rho\Omega^{\dagger}_i),
\label{x}
\end{equation}
are complex numbers, in general. The factor $n$ on R.H.S of (\ref{v}) is to accomodate the unit matrix $\Omega_{0}$ in the chosen %%@
basis. As such $\Omega_{1},...\Omega_{n^{2}-1}$ are traceless.\\
In the particular case of $s=\frac{1}{2},$ the Pauli spin matrices (\ref{f}) satisfy these criteria and are moreover hermitian, so %%@
that the coefficients are real.
Thus, we have
\begin{equation}
\rho=\omega_{0}+\vec{\sigma}\cdot\vec{\omega}
\label{y}
\end{equation}
where $\omega_{0}=\frac{1}{2}Tr{\rho}=\frac{1}{2}$, using (\ref{s}) and we may use (\ref{e}) and (\ref{o}) to identify 
\begin{equation}
\vec{\omega}=\frac{1}{2}Tr({\rho\vec{\sigma}})=\langle\vec{S}\rangle,
\label{z}
\end{equation}
as the classical value of spin $\vec{S}$ for particles with $s=\frac{1}{2}$.\\
It needs to be mentioned that the rows and columns of (\ref{f}) as well as of the spin density matrix $\rho$, usually, are %%@
labelled by the quantum number $m=s,s-1,.....,-s$ in that order. It may be noted that
\begin{equation}
\rho=
\left(
\begin{array}{cc}
	1&0\\
	0&0
\end{array}\right)\;
  or\;
 \rho=
\left(
\begin{array}{cc}
	0&0\\
	0&1
\end{array}\right)\,,
\label{aa}
\end{equation}
if $|\psi\rangle$ itself is $|\frac{1}{2},m\rangle$ with $m=\pm\,\frac{1}{2}$.

\end{section}

\begin{section}{Statistical Matrix or Density Matrix for A Statistical Assembly of Particles with Spin}

 A statistical assembly, like a beam or target used in experiments in physics, is essentially a system of a large number $N$ of %%@
particles, whose wave functions do not have any spatial overlap. We may define the statistical expectation value for spin %%@
$\vec{S}$ in such a system through 
\begin{equation}
E(\langle\vec{S}\rangle)=\frac{1}{N}\sum_{i=1}^{N}\langle\vec{S_i}\rangle=\frac{1}{N}\sum_{i}^{N}Tr({\rho_{i}\vec{S}})=Tr({\rho}
\vec{S})
\label{ab}
\end{equation}
so that the assembly is characterised by what was referred to as the `statistical matrix' or merely as the density matrix for the %%@
statistical assembly
\begin{equation}
\rho=\frac{1}{N}\sum_{i=1}^{N}{\rho_{i}}
\label{ac}
\end{equation}
and (\ref{ab}) may be referred to as the `average expectation value' of $\vec{S}$. The properties (\ref{r}) and (\ref{s}) are %%@
valid, but not necessarily (\ref{t}). The statistical matrix or the density matrix $\rho$ for a statistical assembly may be %%@
defined through (\ref{ac}), in general, for any $s$.\\
 In the particular case of $s=\frac{1}{2}$, for example, we have
\begin{equation}
\rho=\frac{1}{2}[1+\vec{\sigma}\cdot\vec{P}]=
\frac{1}{2}\left(
\begin{array}{cc}
	1+P_{z}&P_{x}-iP_{y}\\
	P_{x}+iP_{y}&1-P_{z}
\end{array}
\right)
\label{ad}
\end{equation}
where the Polarization Vector,\,$\vec{P}$ is given by
\begin{equation}
\vec{P}=Tr({\rho}\vec{\sigma})=2E(\langle\vec{S}\rangle)=\frac{1}{N}\sum_{i=1}^{N}2\,\vec{\omega_{i}}
\label{ae}
\end{equation}
where $\vec{\omega_{i}}$ is given by (\ref{z}) for each particle $i$.\\
Moreover, if we consider $N_{\pm}$ particles to be in the $|\frac{1}{2}\pm\frac{1}{2}\rangle$ states respectively, such that %%@
$N_{+}+N_{-}=N$, we may use (\ref{aa}) to express ${\rho}$ defined by (\ref{ac}) as
\begin{equation}
{\rho}=\frac{1}{N}
\left(
\begin{array}{cc}
	N_{+}&0\\
	0&N_{-}
\end{array}
\right)=
\left(
\begin{array}{cc}
	p_{+\frac{1}{2}}&0\\
	0&p_{-\frac{1}{2}}
\end{array}
\right)
\label{af}
\end{equation}
We may compare (\ref{ad}) and (\ref{af}), if $P_{x}=P_{y}=0$, when we may identify
\begin{equation}
\frac{N_{\pm}}{N}=\frac{1\pm P_{z}}{2}=p_{\pm\frac{1}{2}}
\label{ag}
\end{equation}
as the probabilities $p_m$ assigned to states with $m=\pm \frac{1}{2}$.
Clearly
\begin{equation}
{\rho}^{2}\neq{\rho}
\label{ah}
\end{equation}
for a statistical assembly, except in two extreme cases, viz., either $N_{+}=N$ or $N_{-}=N$, when the statistical assembly is %%@
said to be `pure' and ${\rho}^{2}={\rho}$. If $N_{+}=N_{-}$, the statistical assembly is said to be `unpolarized'. In all other %%@
cases (\ref{ah}) holds and the statistical assembly is said to be `mixed'. Thus
\begin{equation}
-1\leq P_{z}=\frac{{N_{+}-N_{-}}}{N}\leq 1
\label{ai}
\end{equation}
Since $\vec{P}$ is measurable experimentally, ${\rho}$ is determinable empirically and we can take ${\rho}$ given by (\ref{ad}) to %%@
the diagonal form (\ref{af}), with $p_{+{1}/{2}}\geq p_{-{1}/{2}}$, simply by rotating the cartesian coordinate system such that %%@
the $z$-axis after rotation is along $\vec{P}$. \,ie., the $z$-component of $\vec{P}$ after rotation is 
$P=|\vec{P}|=\frac{N_{+}-N_{-}}{N}$\,,with $N_{+}\geq N_{-}$ (since L.H.S is positive definite). The assembly is unpolarized if %%@
$P=0$, pure if $P=1$ and mixed if $0<P<1$. It may also be noted that
\begin{equation}
\sum_m p_m=1=M_0;\hspace{1in} \sum_m m\,p_m=\frac{1}{2}\;P=M_1
\label{aj}
\end{equation}
are the moments of the probability distribution with respect to $m$.
\end{section}

\begin{section}{Statistical Assembly of Particles with Higher Spin}
Treating $m$ (or equivalently $S_{z}$) as a variate and assigning probablities $p_m$ to the states $\left|s,m\right\rangle$, even %%@
when $s>\frac{1}{2}$, leads to a particularly simple category of statistical assemblies referred to as `oriented' \cite{e} and the %%@
z-axis w.r.t which the above states are defined through (\ref{h}) is referred to as the `axis of orientation'. Oriented assemblies %%@
are characterised by moments
\begin{equation}
M_k=\sum_m m^kp_m
\label{ak}
\end{equation}
of order $k$ going upto $k=n-1=2s$.\\
In the particular case of polarised assemblies of particles with spin $s=\frac{1}{2}$, the axis of orientation is clearly %%@
collinear with $\vec{P}$ and since the most general form (\ref{ad}) for $\rho$ is characterised only by $\vec{P}$, the assembly is %%@
always `oriented'.\\
In the case of higher spin $s>\frac{1}{2}$, the general form for ${\rho}$ contains many additional parameters Following Fano %%@
\cite{f}, $\rho$ may be expressed as 
\begin{equation}
\rho=\sum_{l=0}^{2s}\sum_{q=-k}^{k}(-1)^{q}\,t_{q}^{k}\,\tau_{q}^{k}\,,
\label{al}
\end{equation}
where $\tau_{q}^{k}$ are $k^{th}$ degree homogeneous polynomial in $S_x$,\,$S_y$,\,$S_z$. They were referred to originally %%@
\cite{g} as polarized Harmonics and defined by operating the invariant $(\vec{S}.\vec{\nabla})^{k}$ on the solid harmonics ${\cal %%@
Y}_{kq}(\vec{r})=r^k\,{Y}_{kq}(\theta,\varphi)$ where ${Y}_{kq}(\theta,\varphi)$ denote standard Spherical Harmonics, satisfying %%@
the Time Reversal requirement of Wigner \cite{h}.i.e.,
\begin{equation}
\tau^k_q=N_k(\vec{S}.\vec{\nabla})^{k}r^k{Y}_{kq}(\theta,\varphi)\,,
\label{am}
\end{equation}
which are irreducible tensors \cite{i} of rank $k$. The parameters
\begin{equation}
t^k_q=Tr(\rho \tau^k_q)\,,
\label{an}
\end{equation}
in (\ref{al}) are known as Fano Statistical Tensors. The normalization factor $N_k$ in (\ref{am}) depends not only on $k$ but also %%@
on $s$. Instead of the $N_k$ originally chosen by Fano \cite{f} and others \cite{j}, we shall normalize ${\tau_q^k}$ differently %%@
here, so as to satisfy
\begin{equation}
\left\langle sm'\left|{\tau_q^k}\right|sm\right\rangle=C(sks;mqm')[k];\,[k]=\sqrt{2k+1}\,,
\label{ao}
\end{equation}
which is consistent with \cite{k} and the later Madison convention \cite{l} for spin 1 . Making use of the Wigner-Eckart theorem %%@
\cite{m,n,o} and denoting the Clebsch-Gordan Coefficients by $C$, we may identify $\sqrt{2k+1}$ as the reduced matrix element %%@
$\left\langle s\left|\left|{\tau_q^k}\right|\right|s\right\rangle$\\With the above normalization, it follows that $\tau^0_0=1$ and
\begin{equation}
Tr[{\tau_q^k}{\tau_{q'}^{\dagger k'}}]=n{\delta_{kk'}}{\delta_{qq'}}\,,
\label{ap}
\end{equation} 
which is akin to (\ref{v}). Since $q=k,k-1,.....-k$ takes $(2k+1)$ values, it follows that 
\begin{equation}
\sum_{k=1}^{2s}(2k+1)=n^2-1\,,
\label{aq}
\end{equation} 
which coincides exactly with the number of $\Omega_i$, $i=1,....,n^2-1$ required in (\ref{w}). By virtue of Wigner's definition %%@
\cite{h} of the spherical harmonics $Y_{lm}$ the hermitian conjugate ${\tau_q^{\dagger k}}$ of ${\tau_q^k}$ satisfies 
\begin{equation}
{\tau_q^{\dagger k}}=(-1)^q{\tau_{-q}^{k}}\,,
\label{ar}
\end{equation} 
which implies that the complex conjugates $({t_q^k})^*$ of ${t_q^k}$ are related to $t^k_{-q}$ through
\begin{equation}
(t_q^{k})^*=(-1)^q{t_{-q}^{k}}\,,
\label{as}
\end{equation} 
 One may also consider \cite{p} $\frac{1}{2}[{\tau_q^k}+(-1)^{q}{\tau_{-q}^k})]$ and %%@
$\frac{1}{2i}[{\tau_q^k}+(-1)^{q}{\tau_{-q}^k})]$ as independent hermitian $\Omega_i$ for (\ref{w}). Moreover, $t^k_0$ are real %%@
for all $k$. These together with Re\,${t_q^k}$, Im\,${t_q^k}$ for $q>0$ constitute a total of $(n^2-1)$ real parameters, apart %%@
from ${t_0^0}=1$. The operators $\tau^{k}_{q}$ as well as the Fano statistical tensors $t^k_q$, being irreducible tensors of rank %%@
$k$, transform under rotations according to 
\begin{equation}
(t_q^{k})_{II}=\sum_{q=-k}^k D^k_{q'q}(\alpha\beta\gamma)(t_{q'}^{k})_{I}\,,
\label{at}
\end{equation} 
where $(\alpha\beta\gamma)$ denote Euler angles of the rotation from one Cartesian coordinate system $(I)$ to another $(II)$ and %%@
$D^k_{q'q}$ denote elements of the standard rotation matrices \cite{m,n} referred to also as Wigner $D$ functions \cite{o}. In %%@
particular
\begin{equation}
\tau^1_q=\sqrt{\frac{3}{s(s+1)}}S^1_q;\tau^2_q=\sqrt{\frac{2\times3\times5}{s(s+1)(2s-1)(2s+3)}}\,(S^1\otimes S^1)^2_q\,,
\label{au}
\end{equation} 
where 
\begin{equation}
S^1_0=S_z;\hspace{0.5in}S^1_{\pm1}=\mp\frac{1}{\sqrt{2}}(S_x\pm iS_y)
\label{av}
\end{equation} 
are referred to as the spherical components of $\vec{S}$. They  transform, under rotations according to (\ref{at}) and as such %%@
constitute an irreducible tensor of rank $k=1$. The notation 
\begin{equation}
(A^{k_1}\otimes B^{k_2})^k_q=\sum_{q_1}C(k_1k_2k;q_1q_2q)A^{k_1}_{q_1}B^{k_2}_{q_2}\,,
\label{aw}
\end{equation} 
used in (\ref{au}) denotes an irreducible tensor of rank $k$ obtained by combining two irreducible tensors $A^{k_1}_{q_1}$ and %%@
$B^{k_2}_{q_2}$ of ranks $k_1$ and $k_2$ respectively. Defining \cite{k,l} cartesian second rank tensors
\begin{equation}
Q_{\alpha\beta}=\frac{1}{2}[\frac{3}{2}(S_{\alpha}S_{\beta}+S_{\beta}S_{\alpha})-\delta_{\alpha\beta}S^2]=\frac{1}{2}{\cal %%@
P}_{\alpha\beta};\;\;\alpha,\beta=x,y,z\,
\label{ax}
\end{equation} 
which are symmetric and traceless, we may express
\begin{equation}
(S^1\otimes S^1)^2_0=\sqrt{\frac{2}{3}}Q_{zz};(S^1\otimes S^1)^2_{\pm 1}={\pm \frac{2}{3}}(Q_{xz}\pm iQ_{yz};(S^1\otimes %%@
S^1)^2_{\pm 2}={\frac{2}{3}}(Q_{xx}-Q_{yy}\pm iQ_{xy}),
\label{ay}
\end{equation} 
so that the Fano statistical tensor $t^2_q$ may be expressed in terms of the cartesian components
\begin{equation}
{P}_{\alpha\beta}=Tr(\rho {\cal P}_{\alpha\beta})
\label{az}
\end{equation} 
of what is commonly referred to as 'Tensor Polarization', while $t^1_q=P^1_q,$ where $P^1_q$ denote the spherical components of %%@
`vector polarization', $\vec{P}$ of a statistical assembly of particles with any arbitrary spin $s$.
A cartesian coordinate system with its z-axis parallel to $\vec{P}$ may be referred to as the Lakin \cite{q} Frame.If an assembly %%@
with $\vec{P}\neq0$ is `oriented', it goes without saying that the axis of orientation must be collinear with $\vec{P}$. It is %%@
clear that $t_q^k$ w.r.t the axis of orientation, should satisfy
\begin{equation}
t^{k}_{q}=B_{k}\delta_{q_0}\,,
\label{ba}
\end{equation} 
for an `oriented system', from which it follows that the probablities
\begin{equation}
p_m=\frac{1}{2s+1}\sum_{k}C(sks,mom)[k]B_k\,,
\label{bb}
\end{equation} 
where
\begin{equation}
B_k=[s]G_k=\sum_{m}C(sks,mom)[k]p_m\,,
\label{bc}
\end{equation} 
have been referred to as `Orientation Parameters' \cite{r}.\\
The probablities $p_m$\;or the moments $M_k$ of (\ref{ak}) or the $B_k$ or $G_k$ of (\ref{bc}) constitute a set of $(n-1)$ real %%@
and independent parameters. Together with the polar and azimuthal angles $(\theta,\varphi)$ needed to sepcify teh axis of %%@
orientation in space, an `oriented' spin assembly is completely characterised by $(n+1)$ real independent parameters. On the other %%@
hand, we have pointed out below (\ref{as}) that a general analysis of ${\rho}$ shows that one needs a set of $(n^2-1)$\;real %%@
independent parameters to characterise a statistical assembly of particles with spin $s$. Clearly
\begin{equation}
n^2-1\geq n+1\,,
\label{bd}
\end{equation} 
where the equality holds only in the case of $n=2$ or $s=\frac{1}{2}$. This shows that when we consider statistical assemblies of %%@
higher spin particles, an `oriented' system \cite{s} constitutes only a particularly simple case, where the $t^k_q$ are not %%@
independent but have $(n^2-n-2)$ constraints \cite{s}. Oriented systems have cylindrical symmetry with respect to the axis of %%@
orientation; as such $\rho$ assumes the diagonal form $\rho^0$, when the axis of orientation is chosen as the $z$-axis.\\
\end{section}

\begin{section}{Non-Oriented Statistical Assemblies and $SU(n)$ Representation}

 One can envisage statistical assemblies of particles with higher spin $s>\frac{1}{2}$, which are more general. Such assemblies, %%@
with no manifestly cylindrical symmetry, have been named as \cite{t} as `Non-oriented spin systems'; the $\rho$ for these systems %%@
cannot be brought to the diagonal form by any suitable choice of the $z$-axis (i.e., the axis quantization).\\
In such a general case, it is more convenient \cite{t} to choose, ${\Omega_i}$ in (\ref{w}) as the generators \cite{u,v} of the %%@
compact semi-simple Lie group $SU(n)$. This group is characterised by $(n-1)$ diagonal matrices %%@
$H_{\alpha}$,\;${\alpha}=1,....,n-1$ and $(n^2-n)$ off-diagonal matrices $E^{\gamma}_{\alpha\beta}$, ${\beta=\alpha+1,...,n}$ and %%@
$\gamma=1,2$. Labelling the rows and columns by $\mu,\nu=1,...,n$ these matrices are given in terms of their elements 
\begin{eqnarray}
(H_{\alpha})_{\mu\nu}=\left[\frac{n}{\alpha(\alpha+1)}\right]^{\frac{1}{2}}\left\{
\begin{array}{cc}
	\delta_{\mu\nu}& \texttt{if}\;\;\,\mu<\alpha+1\\
	-\alpha\delta_{\mu\nu}& \texttt{if}\;\;\,\mu=\alpha+1\\
	0 & \texttt{if}\;\;\,\mu>\alpha+1
\end{array}
\right.
\label{be}
\end{eqnarray}
\begin{equation}
(E_{\alpha\beta}^1)_{\mu\nu}=\sqrt{\frac{n}{2}}\,{\delta_{\mu\alpha}}{\delta_{\nu\beta}}=(E_{\alpha\beta}^1)_{\nu\mu};\; %%@
{\nu>{\mu}}
\label{bf}
\end{equation}
\begin{equation}
(E_{\alpha\beta}^2)_{\mu\nu}=-i\sqrt{\frac{n}{2}}\,{\delta_{\mu\alpha}}{\delta_{\nu\beta}}=(E_{\alpha\beta}^2)^*_{\nu\mu};\;{\nu>{%%@
\mu}}
\label{bg}
\end{equation}
and are as such hermitian and traceless. They satisfy (\ref{v}) i.e.,
\begin{equation}
Tr(H_{\alpha}H_{\beta})=n{\delta_{\alpha\beta}}\,,
\label{bh}
\end{equation}
\begin{equation}
Tr\left(E_{\alpha\beta}^{(\gamma)} %%@
E_{\alpha'\beta'}^{(\gamma')}\right)=n{\delta_{\alpha\alpha'}}{\delta_{\beta\beta'}}{\delta_{\gamma\gamma'}}\,,
\label{bi}
\end{equation}
\begin{equation}
Tr\left(E_{\alpha\beta}^{(\gamma)} H_{\alpha}\right)=0\,.
\label{bj}
\end{equation}
It may readily be seen in particular that $H_{1}=\sigma_{z},\,E^{1}_{12}=\sigma_{x},\,E^{2}_{12}=\sigma_{y}$ for $n=2$ or %%@
$s=\frac{1}{2}$. In the case of $s=1$ or $n=3$, the $H_{\alpha},\, E_{\alpha\beta}^{(\gamma)}$ are given by the eight Gellmann %%@
matrices[23]. Thus $(H_{\alpha},E_{\alpha\beta}^{(\gamma)})$ together with the unit matrix $\Omega_0=1$ constitute a set of $n^2$ %%@
linearly independent $n\times n$ matrices and we may express 
\begin{equation}
\rho=\frac{1}{n}\left[1+\sum_{\alpha=1}^{n-1}h_{\alpha}H_{\alpha}+\sum_{\alpha=1}^{n-1}\sum_{\beta=\alpha+1}^{n}\sum_{\gamma=1}^{2%%@
%%@
}e_{\alpha\beta}^{(\gamma)}E_{\alpha\beta}^{(\gamma)}\right]\,,
\label{bk}
\end{equation}
where the $(n^2-1)\;SU(n)$parameters are given, using (60) to (62) in (63) by
\begin{equation}
h_\alpha=Tr(\rho H_{\alpha});\;\,e_{\alpha\beta}^{(\gamma)}= Tr\left(\rho E_{\alpha\beta}^{(\gamma)}\right)\,,
\label{bl}
\end{equation}
and they are real. Note that the hermitian  ${\rho}$ can, in general, be brought to the diagonal form $\rho^0$, through a unitary %%@
transformation represented by the matrix, $U$. If $\rho$ is known in the form (38), in terms of the experimentally determined %%@
$t^{k}_{q}$ w.r.t a convinently chosen Laboratory frame ($L$) where the basis states are denoted by $\left|sm\right\rangle_{L}$, %%@
we may equate (38) with (63), provided the basis states $\left|\alpha\right\rangle$ of (63) are identified through %%@
$\left|\alpha\right\rangle=\left|sm\right\rangle_{L}$, where $\alpha=s+1-m$ takes values $1,2...,n$ as $m$ takes values %%@
$m=s,...,-s$. One may then determine the $h_{\alpha}$ in (63) through
\begin{equation}
h_{\alpha}=\frac{1}{\sqrt{n\alpha(\alpha+1)}}\sum_{k}\,[k]\,t_{0}^{k}\left[\sum_{n=0}^{\alpha-1}C(sks;s-n,0,s-n)-\alpha\,C(sks;s-
\alpha,0,s-\alpha)\right]\,,
\label{bm}
\end{equation}
while $e_{\alpha\beta}^{(\gamma)}$ may be determined by equating the real and imaginary parts of 
\begin{equation}
e^{(1)}_{{\alpha},{\beta}}+i\,e^{(2)}_{{\alpha},{\beta}}=\frac{\sqrt{2}}{\sqrt{n}}\sum_{k}(-1)^{q}[k]\,C(sks;s+1-\alpha,-q,s+1- %%@
\beta)\,t_{q}^{k}\,,
\label{bn}
\end{equation} 
The advantage here is that the $SU(n)$ representation (63) of $\rho$ leads, in general, to the diagonal form
\begin{equation}
\rho^0=U{\rho}U^{\dagger}=\frac{1}{n}\left[1+\sum_{\alpha=1}^{n-1}h_{\alpha}^{0}H_{\alpha}\right]\,,
\label{bo}
\end{equation}
which contains only the $H_{\alpha}$,\;${\alpha}=1,....n-1$ and the unit matrix. The eigen states of ${\rho}$ are 
\begin{equation}
\left|\alpha\right\rangle_{0}=\sum_{\alpha'=1}^{n} U_{\alpha \alpha'} \left|\alpha'\right\rangle,\;\alpha=1,...,n\,,
\label{bp}
\end{equation}
while the corresponding eigenvalues are the statistical probabilities
\begin{equation}
p_{\alpha}=\frac{1}{n}\left[1-\sqrt{\frac{n(\alpha-1)}{\alpha}}h^{0}_{\alpha-1}+\sum_{\eta=\alpha}^{n-1}\sqrt{\frac{n}{\eta(\eta %%@
+1)}}\,h_{\alpha}^{0}\right],\;\alpha=1,...,n\,,
\label{bq}
\end{equation}
which add up to 1. The new statistical parameters
\begin{equation}
h_{\alpha}^{0}=\sqrt{\frac{n}{\alpha(\alpha+1)}}\left[\sum_{\eta=1}^{\alpha}p_{\eta}-\alpha %%@
p_{\alpha+1}\right],\;{\alpha=1,...n-1}
\label{br}
\end{equation}
are sufficiently general to describe oriented as well as non-oriented statistical assemblies.\\
In the simplest case of an `oriented' assembly, the $\left|\alpha\right\rangle_{0}$ are identifiable as %%@
$\left|s,m\right\rangle_{0}$ defined w.r.t the axis of orientation, since $U$ can be identified with a rotation and %%@
$U_{\alpha\alpha'}=D^{s}_{m'm}(\varphi_{0},\theta_{0},0)$ where $(\theta_{0},\varphi_{0})$ define the axis of orientation in the %%@
Lab Frame. 
In the case of `non-oriented' assemblies, one cannot identify all the eigen states $\left|\alpha\right\rangle_{0}$ as eigen states %%@
of $S_z$, w.r.t any choice of an axis of quantization, since $U$ cannot in this case be identified with a rotation. All rotations %%@
are unitary, but not all unitary transformations are rotations when $n>2$.\\
The second advantage is that one can envisage a correlated inductive procedure to determine the bounds on the $h_{\alpha}^{0}$ for %%@
any arbitary spin, noting simply that
\begin{equation}
0\leq p_{\alpha}\leq 1\,, 
\label{bs}
\end{equation}
\begin{equation}
\sum_{\alpha=1}^{n} p_{\alpha}=1\,,
\label{bt}
\end{equation}
Setting ${\alpha=n-1}$ in (70) and using (72)gives 
\begin{equation}
h_{n-1}^{0}=\frac{1}{\sqrt{n-1}}(1-np_{n}).
\label{bu}
\end{equation}
If $p_n=0$, we have the upper limit, while $p_{n}=1$ leads to the lower limit for $h_{n-1}^{0}$. Thus,
\begin{equation}
-\sqrt{n-1}\leq h_{n-1}^{0}\leq \frac{1}{\sqrt{n-1}}.
\label{bv}
\end{equation}
Setting $\alpha=n-2$ in (70) and using (72) gives
\begin{equation}
h_{n-2}^{0}=\sqrt{\frac{n}{(n-1)(n-2)}}\left[1-p_n-(n-1)p_{n-1}\right]\,,
\label{bw}
\end{equation}
Clearly, the upper limit is realised if $p_n=p_{n-1}=0$. If $p_n=1$, it is clear that all $p_{\alpha\neq n}=0$, while $p_n=0$ %%@
allows $p_{n-1}$ to have any value in the range $0\leq p_{n-1}\leq 1$. The lower limit is realised, if $p_{n-1}=1$ (and hence %%@
$p_n=0$). Thus
\begin{equation}
-\sqrt{\frac{n(n-2)}{n-1}}\leq h_{n-2}^{0}\leq \sqrt{\frac{n}{(n-1)(n-2)}}\,,
\label{bx}
\end{equation}
it may be noted that $h_{n-2}^{0}=0$, either if $p_{n}=p_{n-1}=\frac{1}{n}$ or if $p_{n}=1$ and hence $p_{n-1}=0$. Setting %%@
$\alpha=n-3$ in (70) and using (72) gives
\begin{equation}
h_{n-3}^{0}=\sqrt{\frac{n}{(n-2)(n-3)}}[1-p_{n}-p_{n-1}-(n-2)p_{n-2}]
\label{by}
\end{equation}
Clearly the upper limit is realised if $p_{n}=p_{n-1}=p_{n-2}=0$,\, while $p_{n-2}=1$ (and hence $p_{n}=p_{n-1}=0$) leads to the %%@
lower limit. Thus,
\begin{equation}
\sqrt{\frac{(n)(n-3)}{(n-2)}}\leq h_{n-3}^{0}\leq \sqrt{\frac{n}{(n-2)(n-3)}}
\label{bz}
\end{equation}
It may be noted that $h_{n-3}^{0}=0$, either if $p_{n}=p_{n-1}=p_{n-2}=\frac{1}{n}$ or if $p_{n}=1$ (and hence $p_{n}=p_{n-2}=0$). %%@
We may then set $\alpha=n-4$ and so on. The results obtained may be summerised as 
\begin{equation}
\sqrt{\frac{n\alpha}{\alpha+1}}\leq h_{\alpha}^{0}\leq \sqrt{\frac{n}{\alpha(\alpha+1)}};\;\alpha=1,...,n-1
\label{ca}
\end{equation}
where the upper and lower limits correspond respectively to $p_{n}=p_{n-1}=...=p_{\alpha+1}=0$ and to $p_{\alpha+1}=1$ and hence %%@
$p_{n}=p_{n-1}=...=p_{\alpha+2}=0$. Thus, (79) is equivalent to the positivity conditions (71) for $p_{\alpha}$, $\alpha=2,...,n$.
Finally the bound $0\leq p_1 \leq 1$ leads to 
\begin{equation}
-1\leq \sum_{\alpha=1}^{n-1}\sqrt{\frac{n}{\alpha(\alpha+1)}}\,h^0_{\alpha} \leq n-1\,,
\label{cb}
\end{equation}
on setting $\alpha=1$ in (67). The structure of the bounds in terms of the $h_{\alpha}^0$, $\alpha=1,...,(n-1)$ is valid, in %%@
general, for a statistical assembly of particles with any spin,\,$s$.\\
More interestingly, we may visualise a real $(n-1)$ dimensional space $\Re^{n-1=2s}$, treating $h_{\alpha}^0$, %%@
$\alpha=1,...,(n-1)$ as orthogonal coordinates. By setting $p_{\alpha}=0$ in (69) for $\alpha=1,...,n$ leads to 
a set ${\{S_{\alpha},{\alpha=1},...,n}\}$ of $(n-2)$ dimensional surfaces in $\Re^{n-1}$ enclosing the origin O, which corresponds %%@
to $h_{\alpha}=0$, $\alpha=1,...,(n-1)$ and $p_{\alpha}=\frac{1}{n}$,\;$\alpha=1,...,n$ as is evident from (69). Therefore %%@
$p_{\alpha}$ increases from $0$ to $\frac{1}{n}$ as we move along the normal to $S_{\alpha}$ towards the origin, O (while %%@
$p_{\alpha}<0$ if we move away from O). Note, moreover, that $p_{\alpha}=1$,\;$\alpha=1,...,n$ constitute $n$ vertices of a %%@
polyhedron inscribed in a hyper sphere of radius $\sqrt{2s}$ in $\Re^{n-1}$. Clearly, the inside of the polyhedron (including its %%@
surface) defines the probability domain, where the centre O corresponds physically to an unpolarized statistical assembly and the %%@
$n$ vertices correspond to the `pure' states, while all other points inside the polyhedron correspond to `mixture' states of the %%@
statistical assembly. 
The above geometrical visualisation of the probability domain in terms of the $h_{\alpha}^0$ is not only elegant, but also valid %%@
for any arbitrary $s$. In the simplest case $s=1$ of higher spins, the polyhedron reduces to an equilateral triangle in $\Re^{2}$, %%@
as shown in Fig 1.
\begin{figure}[h]
	\fbox{\includegraphics[scale=0.3, angle=90]{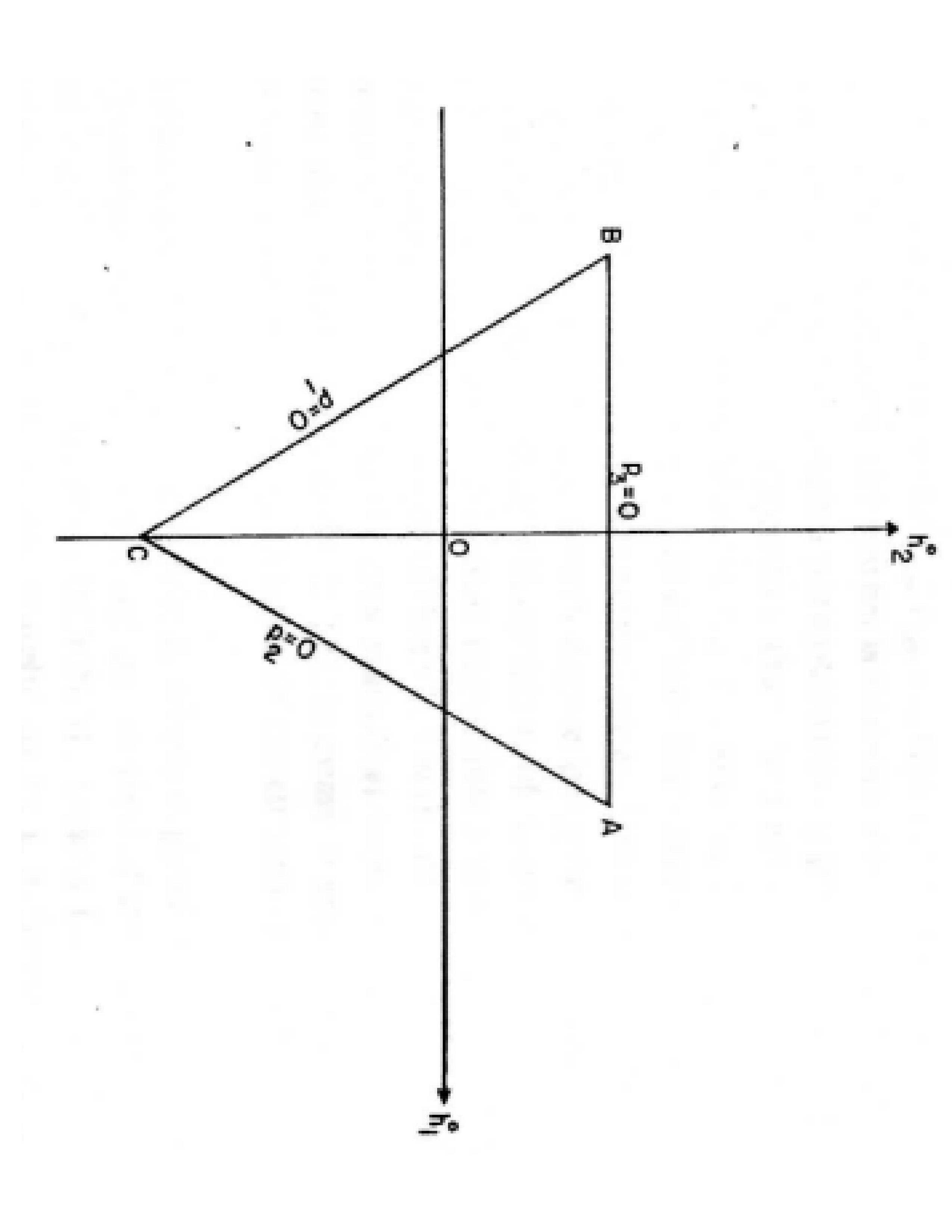}}
	\caption{The allowed region for the new statistical tensors $h_{1}^{0}$ and $h_{2}^{0}$ in the case of spin-1 systems. The %%@
vertices A, B and C of the equilateral triangle corresponding respectively to $p_{1}=1$,\;$p_{2}=1$ and $p_{3}=1$ are given by %%@
$\left(\sqrt{\frac{3}{2}},\,\sqrt{\frac{1}{2}}\right)$,\;$\left(-\sqrt{\frac{3}{2}},\,\sqrt{\frac{1}{2}}\right)$ and %%@
$\left(0,-\sqrt{2}\right)$ in the $h_{1}^{0}-h_{2}^{0}$ plot.}
	\label{fig:1}
\end{figure}
The bounds in terms of the Fano statistical tensors have been discussed by various authors \cite{j,x} in some particular cases %%@
like $s=1,\frac{3}{2}$. For $s=\frac{3}{2}$, the discussion in terms of $h_{\alpha}^{0}$ leads to a tetrahedron [20] in $\Re^{3}$, %%@
shown in Fig 2.
\begin{figure}[h]
\fbox{\includegraphics[scale=0.3, angle=90]{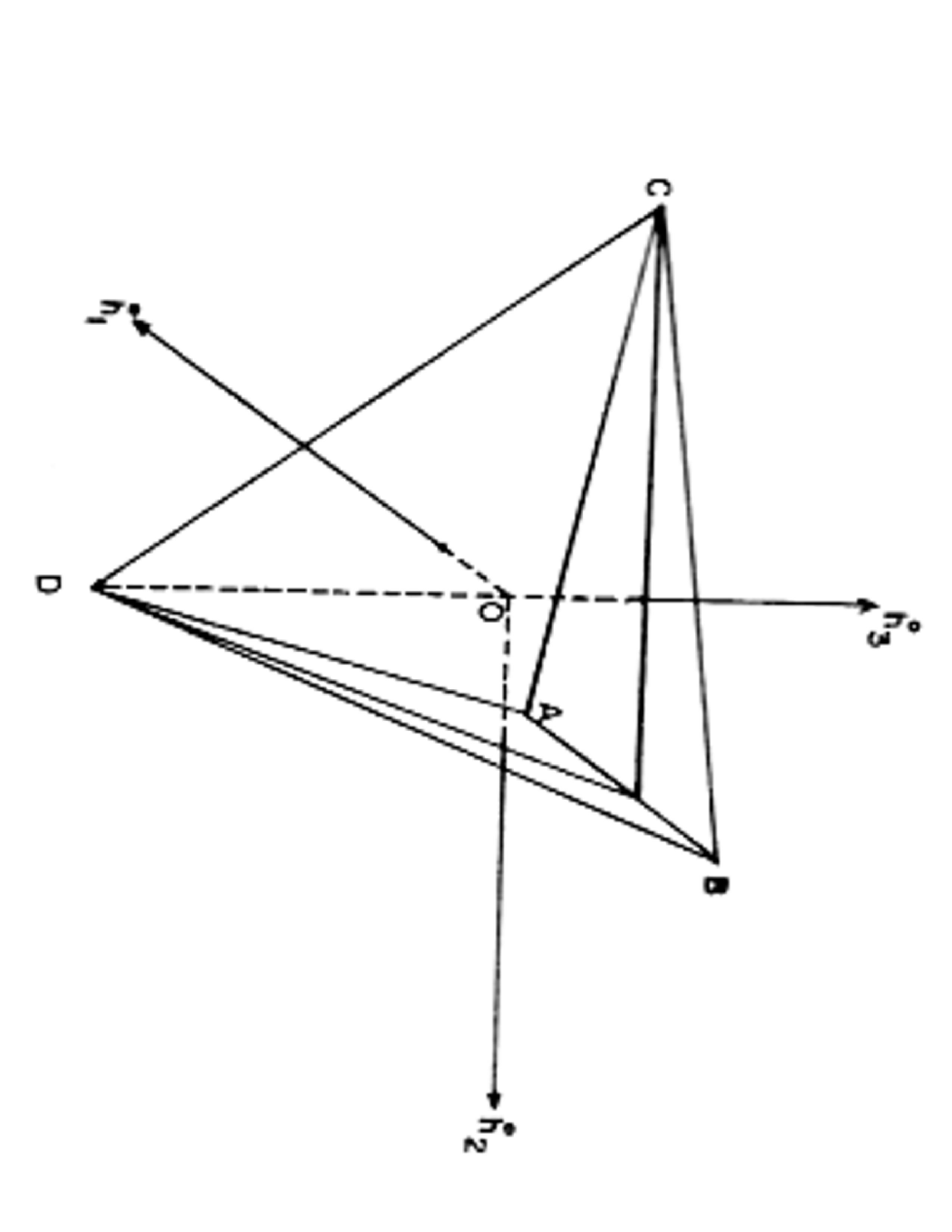}}
\caption{The allowed region for the new statistical tensors $h_{1}^{0}$, $h_{2}^{0}$ and $h_{3}^{0}$ in the case of %%@
spin-$\frac{3}{2}$ systems. The vertices A, B , C and D of the regular tetrahedron corresponding respectively to %%@
$p_{1}=1$,\;$p_{2}=1$,\;$p_{3}=1$ and $p_{4}=1$ are given by %%@
$\left(\sqrt{2},\,\sqrt{\frac{2}{3}},\,\sqrt{\frac{1}{3}}\right)$,\;$\left(-\sqrt{2},\,\sqrt{\frac{2}{3}},\,\sqrt{\frac{1}{3}} %%@
\right)$,\;$\left(0,\,-2\sqrt{\frac{2}{3}},\,\sqrt{\frac{1}{3}}\right)$ and $\left(0,\,0,\,-\sqrt{3}\right)$ in the three %%@
dimensional space spanned by $h_{1}^{0},\;h_{2}^{0}$ and $h_{3}^{0}$.}
\label{fig:2}
\end{figure}
In general, a statistical assembly of particles with spin $s$ may be characterised either in terms of the Fano statistical tensors %%@
\cite{f} or in terms of the $SU(n)$ parameters introduced by Ramachandran and Murthy \cite{u}, which constitute a total of %%@
$(n^2-1)$ variates. In either case, the choice of variates is consistent with Gleason's theorem \cite{y}.

\end{section}
\begin{section}{Multiple Axes of Non-Oriented Statistical Assemblies}
The richness inherent in non-oriented statistical assemblies of particles with higher spin $s>\frac{1}{2}$, may be appreciated %%@
intuitively from the following simple considerations, which is then followed by a formal proof [26].
Any vector $\vec{P}$, such as the one given by (31), may be specified either in terms of its three real components %%@
$P_{x},P_{y},P_{z}$ with respect to a given cartesian coordinate system (say Lab) or equivalently by its spherical components.
\begin{equation}
P_{q}^{1}=P\sqrt{\frac{4\pi}{3}}Y_{1q}(\theta,\varphi);\;q=\pm1,0
\label{cc}
\end{equation}
in terms of its magnitude $P=\left|\vec{P}\right|$ and direction $\hat{P}=\frac{\vec{P}}{P}$, specified by the angles $(\theta, %%@
\varphi)$. Since $P$ is invariant and $P^{1}_{q}$ transform under rotations according to (46), we may envisage an irreducible %%@
tensor
\begin{equation}
\hat{P_{0}^{1}}=cos\theta; \hat{P_{\pm}^{1}}=\frac{1}{\sqrt{2}}sin\theta e^{\pm i\varphi}
\label{cd}
\end{equation}
of rank 1. Clearly, there exists a frame $F$ in which $\hat{P}$ is collinear with the $z$-axis i.e., where $\hat{P}_{\pm}^{1}=0$. %%@
We may next envisage an irreducible tensor
\begin{equation}
t_{q}^{2}(1,2)=\left(\hat{P}(1)\otimes \hat{P}(2)\right)^{2}_{q}
\label{ce}
\end{equation}
of rank 2, constructed out of two given vectors $\vec{P}(1)$ and $\vec{P}(2)$ using (49). Since (83) implies, due to %%@
$C(112;\pm1\pm1\pm2)=1$, that $t^{2}_{\pm2}(1,2)=P_{\pm}^{1}(1)P_{\pm}^{1}(2)$, it follows that there exist two frames $F_{1}$ %%@
$F_{2}$ in which $t_{\pm2}^{2}=0$. Continuing this construction using (49), we may envisage an irreducible tensor 
\begin{equation}
t^{k}_{q}(1,...,k)=\left(...((\hat{P}(1)\otimes \hat{P}(2))^{2}\otimes \hat{P}(3))^{3}\otimes ...\otimes \hat{P}(k)\right)^{k}_{q}
\label{cf}
\end{equation}
of any rank $k$ such that $t_{\pm k}^{k}=0$ in $k$ frames $F_{1}, F_{2},...,F_{k}$. We may now demonstrate, using (46) and %%@
Wigner's formula \cite{i,m,o} for $D_{q}^{k}(\alpha\beta\gamma)$, that any arbitary irreducible tensor $t_{q}^{k}$ of rank $k$ %%@
must indeed be of the form
\begin{equation}
t^{k}_{q}=R_{k} t^{k}_{q}(1,...k)\,,
\label{cg}
\end{equation}
where $R_{k}$ is an arbitary scalar or real number (like the magnitude $P$ in $\vec{P}=P\hat{P}$) which is invariant with respect %%@
to rotations.
Given any $t^{k}_{q}$ in a frame $F_{I}$(say Lab), we pose the question: Is there a frame $F_{II}$ w.r.t which $\left(t^{k}_{\pm %%@
k}\right)_{II}=0?$ i.e.,
\begin{equation}
\left(t^{k}_{\pm k}\right)_{II}=0=\sum_{q=-k}^{k}D_{q,\pm k}^{k}(\varphi,\,\theta,\,\psi)t_{q}^{k}\,,
\label{ch}
\end{equation}
where we use $(\varphi,\,\theta,\,\psi)=(\alpha,\,\beta,\,\gamma)$ for the Euler angles in (86), so that $(\theta,\varphi)$ %%@
denotes the direction of the $z$-axis of $F_{II}$ with respect to Lab frame $F_{I}$, where the $(2k+1)$ complex numbers %%@
$t_{q}^{k}$,\;$q=k,...,-k$ specify the given irreducible tensor of rank $k$. Using Wigner's formula \cite{i,m,o}, we may write
\begin{eqnarray}
D_{q,\pm k}^{k}(\varphi\theta\psi)=(a_{\pm})^{2k}\,\left(
\begin{array}{c}
	2k\\
	k+q
\end{array}
\right)^{1/2}\,Z_{\pm}^{k+q}
\label{ci}
\end{eqnarray}
and take $a_{\pm}$ given by 
\begin{equation}
a_{+}=sin\frac{\theta}{2}e^{i(\varphi-\psi)/2};\;a_{-}=cos\frac{\theta}{2}e^{i(\varphi+\psi)/2}
\label{cj}
\end{equation}
out of the summation w.r.t $q$ in (86). The complex variable $Z$ is given by either 
\begin{equation}
Z_{+}=cot\frac{\theta}{2}e^{i(\varphi)}\;\;\texttt{or}\;\;Z_{-}=-tan\frac{\theta}{2}e^{i(\varphi)}
\label{ck}
\end{equation}
so that (86) implies that
\begin{equation}
a_{\pm}=0
\label{cl}
\end{equation}
or
\begin{equation}
\sum_{q=-k}^{k}C_{q}Z^{k+q}=0
\label{cm}
\end{equation}
The polynomial equation (91) of degree $2k$ with coefficients
\begin{eqnarray}
C_{q}=\left(
\begin{array}{c}
	2k\\
	k+q
\end{array}
\right)^{1/2}\,t_{q}^{k}
\label{cn}
\end{eqnarray}
yields $2k$ solutions for the complex variable,$Z$. These correspond to $2k$ directions %%@
$(\theta_{1},\varphi_{1}),....(\theta_{2k},\varphi_{2k})$ for the $z$-axis of $F_{II}$ w.r.t the Lab frame (where the $t_{q}^{k}$ %%@
are given). We may note from (89) that if $(\theta,\varphi)$ is a solution, then $(\pi-\theta,\pi+\varphi)$ is also a solution. %%@
This corresponds to the inversion of the $z$-axis of $F_{II}$. It is clear that (90) implies the existence of a pair of solutions, %%@
corresponding to $z$-axis of $F_{II}$ being either parallel or antiparallel with the Lab $z$-axis itself. It is worth noting that %%@
$\vec{P}$ in (31) is an axial vector or pseudo- vector, whose components $P_{x},P_{y},P_{z}$ remain unaltered under an inversion %%@
of the coordinate system (a co-ordinate transformation, which is often referred to in physics as Parity). We therefore conclude %%@
that $\vec{P}(1),...,\vec{P}(k)$ considered above must also be axial vectors and that $t_{q}^{k}$ must be of the form (85), which %%@
is specified by a set of $k$ axial vectors or Axes. Thus, the  $(2k+1)$ real degrees of freedom associated with any irreducible %%@
tensor $t_{q}^{k}$ of rank $k$ may be identified with $k$ axes and one real number $R_{k}$, which is essentially a strength factor %%@
(akin to $P=|\vec{P}|$) which is invariant under rotations and inversion. Since the density matrix $\rho$ is specified in (38) by %%@
$t_{q}^{k}$ with $k=1,...,2s$ the $(n^{2}-1)$ parameters characterising $\rho$ may be identified with 
\begin{equation}
N_{A}=\sum_{k=1}^{2s}k=s(2s+1)=sn
\label{co}
\end{equation}
axes and $2s=n-1$ real numbers $R_{k}$.\\
In particular, therefore, a statistical assembly of particles with spin $s=\frac{1}{2}$ is characterised by only one axis viz., %%@
$\hat{P}$ and one scalar or real number $P$, as we have already seen in Sec. V. In the case of $s=1$, the statistical assembly is %%@
characterised by 3 axes and 2 scalars (or real numbers), which specify respectively the strengths of the vector and tensor %%@
polarizations. The number $N_{A}$ of axes increases with $s$. For $s=\frac{3}{2}$, for example, $N_{A}=6$ , while $N_{A}=10$ for %%@
$s=2$ and so on.
A statistical assembly is `oriented', only when all these $N_{A}$ axes collapse into 1. In the particular case of $s=\frac{1}{2}$, %%@
the statistical assembly can only be `oriented', since $N_{A}=1$.
\end{section}

\begin{section}{Particular case of a Non-Oriented Statistical Assembly of Particles with spin $s=1$}
It is perhaps appropriate that we take a closer look at the simplest case $s=1$ of a statistical assembly of higher spin %%@
particles. In general, the density matrix $\rho$, in this case may be written explicitly as 
\begin{equation}
{\rho}=\frac{1}{3}
\left[
\begin{array}{ccc}
1+\sqrt{\frac{3}{2}}\,t_{0}^{1}+\frac{1}{\sqrt{2}}\,t_{0}^{2}&\;\;\;\sqrt{\frac{3}{2}}\,(t_{-1}^{1}+t_{-1}^{2})&\;\;\;\sqrt{3}\, %%@
t_{-2}^{2}\\
	\sqrt{\frac{3}{2}}\,(t_{1}^{1}+t_{1}^{2})&\;\;\;1-\sqrt{2}\,t_{0}^{2}&\;\;\;\sqrt{\frac{3}{2}}\,(t_{-1}^{1}-t_{-1}^{2})\\
	\sqrt{3}\,t_{2}^{2}&\;\;\;-\sqrt{\frac{3}{2}}\,(t_{1}^{1}-t_{1}^{2})&\;\;\;\;1-\sqrt{\frac{3}{2}}\,t_{0}^{1}+{\frac{1}{\sqrt{2%%@
}}}t_{0}^{2}
\end{array}
\right]
\label{cp}
\end{equation}
where $t_{q}^{1}=P \hat{P}^{1}_{q}$ as in (81) and (82) and $t_{q}^{2}=R\left(\hat{P}(1)\otimes \hat{P}(2)\right)^{2}_{q}$ as in %%@
(83),(85). The assembly is thus characterised by three axes $\hat{P}, \hat{P}(1), \hat{P}(2)$ and two scalars Viz, P and R, which %%@
represent respectively the strengths of the vector and tensor polarizations. If $P\neq0$, $R\neq0$, the assembly is non-oriented, %%@
except when all the three axes happen to be collinear. Even if we do away with one axis by setting $P=0$, the statistical assembly %%@
in such a case is said to be `aligned', but it is still non-oriented so long as $\hat{P}(1)$ and $\hat{P}(2)$ are distinct. The  %%@
$t_{q}^{2}$ may explicitly be written as 
\begin{equation}
t^{2}_{0}=\frac{R}{\sqrt{6}}[2\,cos{\theta_{1}}cos{\theta_{2}}-sin{\theta_{1}}sin{\theta_{2}}cos({\varphi_{1}}-{\varphi_{1}})]
\label{cq}
\end{equation}
\begin{equation}
t^{2}_{\pm1}=\mp\frac{R}{2}[(cos{\theta_{1}}sin{\theta_{2}}cos{\theta_{2}}+cos{\theta_{2}}sin{\theta_{1}}cos{\varphi_{1}}]]\pm %%@
i(cos{\theta_{1}}sin{\theta_{2}}sin{\varphi_{2}}+cos{\theta_{2}}sin{\theta_{1}}sin{\varphi_{1}})]
\label{cr}
\end{equation}
\begin{equation}
t^{2}_{\pm2}=\frac{R}{2}sin{\theta_{1}}sin{\theta_{2}}[cos({\varphi_{1}}+{\varphi_{1}})\pm i sin({\varphi_{1}}-{\varphi_{1}})
\label{cs}
\end{equation}
in terms of $R,\;\hat{P}(1),\;\hat{P}(2)$. The Fano statistical tensors may also be expressed in terms of the cartesian components %%@
$P_{\alpha\beta}$\;;$\alpha,\,\beta=x,y,z$ given by (52) as 
\begin{equation}
t^{2}_{0}=\frac{1}{\sqrt{2}}P_{zz}
\label{ct}
\end{equation}
\begin{equation}
t^{2}_{\pm1}=\mp\frac{1}{\sqrt{3}}\left(P_{xz}+iP_{yz}\right)
\label{cu}
\end{equation}
\begin{equation}
t^{2}_{\pm 2}=\frac{1}{2\sqrt{3}}\left(P_{xx}-P_{yy}\pm 2iP_{xy}\right)
\label{cv}
\end{equation}
Since $P_{\alpha\beta}$ constitute a traceless symmetric second rank Cartesian tensor, one can bring it to the diagonal form %%@
$P_{\alpha\alpha}^{0}\delta_{\alpha\beta}$; $\alpha,\beta=x_{A},y_{A},z_{A}$, through a rotation of the coordinate system. The %%@
co-ordinate system $\left(x_{A},y_{A},z_{A}\right)$ has been referred to as the Principal Axes of Alignment Frame \cite{aa} or %%@
PAAF for short, where $P_{xy}=P_{yz}=P_{xz}=0$ and consequently $t_{\pm 1}^{2}=0$ and $t_{2}^{2}=t_{-2}^{2}$. Clearly, the two %%@
vectors $\vec{P}(1)$ and $\vec{P}(2)$ define a plane. Comparing (98) to (100) respectively with (95) to (97), reveals \cite{ab} %%@
that the plane containing $\vec{P}(1)$ and $\vec{P}(2)$ must be either $z_{A}-x_{A}$ or $x_{A}-y_{A}$ or $y_{A}-z_{A}$ planes %%@
where the principal axes $y_{A}$ or $z_{A}$ or $x_{A}$ act as a bisector of the angle between $\vec{P}(1)$ and $\vec{P}(2)$. These %%@
cases correspond respectively to the ratio 
\begin{equation}
r=\left(\frac{t_{0}^{2}}{t_{2}^{2}}\right)_{PAAF}
\label{cw}
\end{equation}
being in the intervals $-\infty$ to $-\sqrt{\frac{2}{3}}$, or $-\sqrt{\frac{2}{3}}$ to $\sqrt{\frac{2}{3}}$ or %%@
$\sqrt{\frac{2}{3}}$ to $\infty$. The density matrix (94) with $t_{q}^{1}=0$;\,$t_{\pm1}=0$ and $t_{2}^{2}=t_{-2}^{2}$ in PAAF may %%@
be diagonalized through a Unitary transformation to yield the eigen values
\begin{equation}
p_{x}=\frac{1}{3}\left(1+\frac{1}{\sqrt{2}}t_{0}^{2}+\sqrt{3}t_{2}^{2}\right)
\label{cx}
\end{equation}
\begin{equation}
p_{y}=\frac{1}{3}\left(1+\frac{1}{\sqrt{2}}t_{0}^{2}-\sqrt{3}t_{2}^{2}\right)
\label{cy}
\end{equation}
\begin{equation}
p_{z}=\frac{1}{3}\left(1-\sqrt{2}t_{0}^{2}\right)
\label{cz}
\end{equation}
the corresponding eigen states of $\rho$ being 
\begin{equation}
\left|x_{A}\right\rangle=\frac{1}{\sqrt{2}}\left(\left|1,-1\right\rangle_{A}-\left|1,1\right\rangle_{A}\right)
\label{da}
\end{equation}
\begin{equation}
\left|y_{A}\right\rangle=\frac{1}{\sqrt{2}}\left(\left|1,-1\right\rangle_{A}+\left|1,1\right\rangle_{A}\right)
\label{db}
\end{equation}
\begin{equation}
\left|z_{A}\right\rangle=\left|1,0\right\rangle_{A}
\label{dc}
\end{equation}
where the eqns (\ref{da}), (\ref{db}) and (\ref{dc}) as well as the states $\left|1,m\right\rangle_{A}$ are w.r.t. PAAF. %%@
Interestingly, they satisfy
\begin{equation}
S_{\alpha}\left|\alpha\right\rangle=0, \,\alpha=x_{A},y_{A},z_{A}
\label{dd}
\end{equation}
These three states do not correspond to eigen states of a single operator with different eigen values, but are eigen states of 3 %%@
different operators with the same eigen value viz., zero. Labeling the rows and columns by the states defined by (105) to (108), %%@
the diagonal form $\rho^{0}$ of $\rho$ is given by
\begin{equation}
{\rho}=\frac{1}{3}
\left[
\begin{array}{ccc}
1+\sqrt{\frac{3}{2}}\,h_{1}^{0}+\frac{1}{\sqrt{2}}\,h_{2}^{0}&\;\;\;0&\;\;\;0\\
0&\;\;\;1-\sqrt{\frac{3}{2}}\,h_{1}^{0}+{\frac{1}{\sqrt{2}}}h_{2}^{0}&\;\;\;0\\
0&\;\;\;0&\;\;\;\;1-\sqrt{2}\,h_{2}^{0}
\end{array}
\right]
\label{df}
\end{equation}
in terms of the $SU(n)$ parameters, so that we readily identify
\begin{equation}
h_{1}^{0}=\sqrt{2}t_{2}^{2};\;\;h_{2}^{0}=t_{0}^{2}
\label{dg}
\end{equation}
in terms of the $t_{q}^{k}$ in PAAF. It is worth noting  that we cannot identify here a single variable like $m$, which was %%@
considered as a variate in  the case of oriented assemblies.
\end{section}

\begin{section}{Higher spin particles in final state}

Statistical assemblies of particles with spin, $s$ are also produced in experiments when particles collide with each other or with %%@
nuclei. We may, in general, consider a collision between $a$ and $b$ producing $c$ and $d$, where $a, b, c, d$ have spins $s_a, %%@
s_b, s_c, s_d$ respectively and describe the physical process, in quantum theory, by
\begin{equation}
M_{m_{c}m_{d};m_{a}m_{b}}\equiv\left\langle s_{c}m_{c};s_{d}m_{d}\left|M\right|s_{a}m_{a};s_{b}m_{b}\right\rangle\,,
\label{dh}
\end{equation}
which constitute the elements of a matrix $M$ (with $n_{c}n_{d}$ rows and $n_{a}n_{b}$ columns) at a conserved relativistic energy %%@
$E$ (which includes all the masses as well) in a Lorentz frame referred to as c.m. frame (where the conserved total momentum is %%@
zero). If $\rho^{a}$ (with $n_{a}$ rows and $n_{a}$ columns) and $\rho^{b}$ (with $n_{b}$ rows and $n_{b}$ columns) denote states %%@
of polarization of the beam and target employed initially in the experiment, the final state polarization is described by 
\begin{equation}
\rho^{f}=M\rho^{i}M^{\dagger}\,,
\label{di}
\end{equation}
where $\rho^{i}$ is the direct product of $\rho^{a}$ and $\rho^{b}$. Due to quantum entanglement, $\rho^f$ is not, in general, %%@
expressible as a direct product of a $\rho^{c}$ and a $\rho^{d}$. However, if no observations are made on the spin state of say, %%@
$d$, one can define the spin state of $c$ by  $\rho^{c}$, whose elements are given by
\begin{equation}
\rho^{c}_{m_{c}m_{c'}}=\sum_{m_{d}}\rho^{f}_{m_{c}m_{d};m_{c'}m_{d}}\,.
\label{dj}
\end{equation}
After the high energy electron scattering experiments revealed the electromagnetic structure of the nuclei and  even of the %%@
nucleons themselves, meson factories came up to study their hadronic structure. The meson beams are generated usually through %%@
photoproduction or electro-production. Although these experiments were not highly successful in revealing the hadronic structure %%@
due to the absence of a theory (for meson scattering) which can make precise quantitative predictions like Quantum Electro %%@
Dynamics (which applies in the case of electron and muon scattering). However, improved experimental facilities to study photo- %%@
and electro-production of pseudoscalar and vector mesons have become available, with the advent of the new generation of electron %%@
accelerators at JLab, MIT, BNL in USA, ELSA at Bonn and MAMI at Mainz in Germany, ESRF at Grenoble in France and Spring8 at Osaka %%@
in Japan, with energies going upto 8 GeV. It is, therefore, of interest to allude to photoproduction of $\omega$ mesons (with spin %%@
$s=1$), since a theoretical formalism has been developed recently \cite{ac}, leading to the elegant derivation of formulae for all %%@
spin observables associated with the photoproduction of mesons with arbitrary spin-parity . This new theoretical formalism is %%@
based on the observation that the 'ket' and 'bra' spin states in quantum theory
\begin{equation}
\left|sm\right\rangle=K_{m}^{s};\;\; \left\langle sm\right|=i^{2m} B^{s}_{-m}
\label{dk}
\end{equation}
transform, under rotations, like irreducible tensors $K_{m}^{s}$ and $B^{s}_{m}$ of rank, $s$ so that one can describe [28] a %%@
transistion from an initial state with spin $s_{i}$ to a final state with spin $s_f$ through irreducible tensor operators
\begin{equation}
S_{\mu}^{\lambda}(s_{f},s_{i})=i^{2si}[s_f]\left(K^{s_f}\otimes B^{s_i}\right)^{\lambda}_{\mu}
\label{dl}
\end{equation}
of rank $\lambda=|s_i-s_f|,....|s_i+s_f|$. Without going into the details, it is interesting to point out that the $\rho^{\omega}$ %%@
of the $\omega$ meson, which is photo or electro produced on protons, can be changed by suitable initial spin preparations of the %%@
photon or electron beam and the proton target. Moreover, it is worth being pointed out that the photon, whose spin is 1, has only %%@
two states $\left|1,\pm1\right\rangle$ w.r.t its direction of propagation and these correspond to the right and left circular %%@
states of polarization. The absence of the $\left|1,0\right\rangle$ state could be traced (through an analysis of the %%@
inhomogeneous Lorentz group, referred to also as Poincare group) to the fact that the mass of the photon is zero. This analysis %%@
has also shown that helicity, (which is the component of the spin $\vec{S}$ of a particle along the direction of its momentum and %%@
as such is invariant) takes over the role of the magnetic quantum number $m$ in a relativistic theory. The theoretical framework %%@
in \cite{ac} takes all these aspects into consideration and shows for the first time how the reaction matrix elements (111) may be %%@
analysed in terms of electric and magnetic multipole amplitudes for the production of mesons with arbitrary non-zero spin $s$. The %%@
vector and tensor \cite{ae} polarizations of the meson are measurable respectively by making use of its decay modes %%@
$\omega\rightarrow\pi^{0}+{\gamma}$ and $\omega\rightarrow3\pi$.  It is encouraging to note that the dominant decay mode %%@
$\omega\rightarrow3\pi$ is already being used and experimental work at WASA \cite{af} is expected to facilitate the use  of even %%@
$\omega\rightarrow\pi^{0}+{\gamma}$, with the smaller branching ratio of $8.92\%$. The decay of polarised Delta \cite{ag} with %%@
spin 3/2 is another interesting example of current interest in the context of neutral pion production in proton-proton collisions %%@
\cite{ah}\\
\end{section}
\vspace{-1cm}
\begin{acknowledgments}
I am grateful to Dr.B.Ramachandran for encouragement. I thank Mr. Sujith Thomas for patiently preparing the Latex version %%@
originally. My thanks are  also due to  Dr. B. M. Sankarshan  and Dr S. P. Shilpashree for assisting  me in the preparation of the %%@
present manuscript.
\end{acknowledgments}

\end{document}